\input phyzzx.tex
\tolerance=1000
\voffset=-0.0cm
\hoffset=0.7cm
\sequentialequations

\def\t1{{\tilde 1}}

\def\t{\theta}


\REF{\HAW}{Commun. Math. Phys. {\bf 43} (1975) 199; Phys. Rev. {\bf D14} (1976) 2460.}
\REF{\ADS}{J. Maldacena, Adv. Theor. Math. Phys. {\bf 2} (1998) 231, [arXiv:hep-th/9711200]; S. Gubser, I. Klebanov and A. Polyakov, Phys. Lett. {\bf B428} (1998) 105, [arXiv:hep-th/9802109]; E. Witten, Adv. Theor. Math. Phys. {\bf 2} (1998) 253, [arXiv:hep-th/9802150].}
\REF{\QC}{see for example, S. Lloyd, [arXiv:quant-ph/0501135]; P. Hayden and J. Preskill, JHEP {\bf 0709} (2007) 120, [arXiv:0708.4025]; L. Susskind, [arXiv:1810.11563]; A. Almheiri, X. Dong and D. Harlow, JHEP {\bf 04} (2015) 163, [arXiv:1411.7041].}
\REF{\ZHA}{Y. Zhao, [arXiv:1912.00909].}
\REF{\PIR}{L. Piroli, C. Sunderhauf and X. Qi, JHEP {\bf 2020} (2020) 63; [arXiv:2002.09236].}
\REF{\MAT}{S. Mathur, Class. Quant. Grav. {\bf 26} (2009) 224001, [arXiv:hep-th/0909.1038].}
\REF{\PAG}{D. Page, Phys. Rev. Lett. {\bf71} (1993) 1291, [arXiv:gr-qc/9305007]; Phys. Rev. Lett. {\bf 71} (1993) 3743,
[arXiv:gr-qc/9306083].}
\REF{\HM}{G. T. Horowitz and J. M. Maldacena, JHEP {\bf 0402} (2004) 008, [arXiv:hep-th/0310281].}
\REF{\NIE}{M. A. Nielsen and I. L. Chuang, Quantum Computation and Quantum Information, Cambridge University Press, 2010.}
\REF{\PRE}{G. Auletta, M. Fortunato and G. Parisi, Quantum Mechanics, Cambridge Univerity Press 2009.}
\REF{\TEL}{G. H. Bennett, G. Brassard, C. Crpeau, R. Jozsa, A. Peres and W. K. Wooters, Phys. Rev. Lett. {\bf 70} (1993) 1895.}
\REF{\TIM}{S. Lloyd and J. Preskill, JHEP {\bf 08} (2014) 126, [arXiv:1308.4209]; R. Bousso and D. Stanford, Phys. Rev. {\bf D89}
(2014) 044038, [arXiv:1310.7457].}
\REF{\GOT}{D. Gottesman and J. Preskill, JHEP {\bf 0403} (2004) 026, [arXiv:hep-th/0311269]; S. Lloyd, [arXiv:quant-phys/0406205].}
\REF{\FIRE}{A. Almheiri, D. Marolf, J. Polchinski and J. Sully, JHEP {\bf 1302} (2013) 062, [arXiv:1207.3123];
A. Almheiri, D. Marolf, J. Polchinski, D. Stanford and J. Sully, JHEP {\bf 1309} (2013) 018, [arXiv:1304.6483].}
\REF{\EDI}{E. Halyo in preparation.}
\REF{\ER}{J. M. Maldacena and L. Susskind, Fortsch. Phys. {\bf 61} (2013) 781, [arXiv:1306.0533].} 
\REF{\ALM}{A. Almheiri, T. Hartman, J. Maldacena, E. Shaghoulian and A. Tajdini, [arXiv:2006.06872] and references therein.}

\singlespace
\pagenumber=0
\normalspace
\medskip
\bigskip
\titlestyle{\bf{A Model for Unitary Black Hole Evaporation}}
\smallskip
\author{ Edi Halyo{\footnote*{email: edi.halyo@gmail.com}}}
\smallskip
\centerline {Department of Physics} 
\centerline{Stanford University} 
\centerline {Stanford, CA 94305}
\smallskip
\vskip 2 cm
\titlestyle{\bf Abstract}
We describe a model for unitary black hole evaporation with no information loss in terms of a quantum computation. We assume that there is a fine tuned interaction between the qubits of the black hole Bell states which is the inverse of premeasurement. Evaporation is unitary due to a projection on the black hole state at the end of each computation whereas information comes out of the black hole by teleportation. The model requires only four operations; the qubit interactions, the Hadamard and CNOT gates and the projection which is nonunitary. The model is a concrete quantum computation that realizes the final black hole state idea with some modifications.

\singlespace
\vskip 0.5cm
\endpage
\normalspace

\centerline{\bf 1. Introduction}
\medskip

Quantum mechanics is a unitary theory that preserves information. However, since Hawking's discovery of black hole radiation
it became apparent that black hole evaporation, at least in semiclassical gravity, does not seem to be a unitary process[\HAW]. On the other hand, based on the AdS/CFT correspondence[\ADS], we expect all physics, including black hole evaporation to be unitary. The celebrated black hole information problem is simply this discrepancy between the semiclassical results and the expected unitarity. 

The black hole information problem is actually two closely related problems. The first one is the violation of unitarity in black hole evaporation. Assuming that the black hole is formed in a pure state, the final state of black hole evaporation 
in semiclassical gravity is Hawking radiation in a mixed state. This clearly violates unitarity of quantum mechanics. The second problem is how the information in the black hole comes out in Hawking radiation. For an observer outside the black hole, the matter that forms it asymptotically reaches the (stretched) horizon and thermalizes. Thus, the information is not destroyed but remains on the horizon albeit in a completely scrambled way. It is not clear how this information comes out in Hawking radiation. In principle either problem can be solved separately even though their solutions are expected to be related.

Recently, it was realized that, in many respects, black holes behave like quantum computers[\QC]. Assuming this, we describe black hole evaporation in terms of the operations of a quantum computer{\footnote1{For a different attempt in terms of random quantum circuits see [\ZHA,\PIR].}}. Therefore, in the following
we use the terms black hole/quantum computer and particle/qubit interchangeably.
In this picture, the initial state of the black hole and its near horizon region is the input. As usual, we assume that this is a pure state of the relevant qubits. The final state of the black hole and Hawking radiation constitutes the output. We would like to obtain an output in a pure state that contains all the information of the input.
Everything that happens in between the input and output corresponds to operations made by the quantum computer which constitutes the computation. 
We find that, in order to obtain unitary black hole evaporation, we need to resort to a projection that selects the black hole state
at the end of each computation. This violates unitarity of quantum mechanics in the black hole. Therefore, we assume that quantum mechanics may be violated in the black hole as long as physics outside the black hole, and especially black hole evaporation, is unitary. Thus, we actually consider the black hole to be a``modified" quantum computer with one nonunitary operation.

The simplest description of the information problem is based on the fact that Hawking radiation follows the creation of maximally entangled pairs of particles near the horizon[\MAT]. Hawking radiation occurs when one particle falls into the black hole and the other one escapes to infinity. The entanglement entropy of the pair is
$S_{ent}=ln2$. Therefore, each pair creation adds $ln 2$ to the entanglement entropy of either the black hole or Hawking radiation. Assuming that $N$ Hawking particles are emitted over the lifetime of the black hole, the radiation has an entanglement entropy of $S_{ent} \sim N ln2$ when the black hole evaporates completely. This means that Hawking radiation 
is in a mixed state which violates unitarity.

The simple analysis above seems to be incomplete since it does not address how the thermodynamic (coarse grained) entropy of the black hole, $S_{bh}$, decreases when a Hawking particle is emitted. Semiclassical gravity dictates that the mass and the entropy of the black hole decrease by $1/R_{bh}$ and $ln2$ respectively. How this happens is not clear in the above picture since the infalling Hawking particle seems to increase $S_{bh}$ by $ln 2$ rather than decrease it. 
On the other hand, the mass of the black hole decreases during Hawking radiation since the infalling particle has negative energy
inside the horizon. Then, by the first law of thermodynamics, the infalling particle should also carry negative entropy. We conclude that, for a complete description of black hole evaporation, there must be two types of qubits: ``positive" qubits with positive energy and entropy and ``negative" qubits with negative energy and entropy. The former are the usual qubits that contribute to the mass and entropy of the black hole. An example of the latter are the infalling particles. Thus, $S_{bh}$ decreases when a Hawking particle is emitted since the infalling particle carries a negative entropy of $-ln2$.
The existence of positive and negative qubits raises the possibility of having positive--negative qubit pairs with vanishing energy and entropy. Below, we will see that these qubit pairs play a crucial role in the description of unitary black hole evaporation that preserves information. In particular, we will find out that the 
positive--negative qubit pairs describe the vacua at different stages of evaporation. 

The existence of negative qubits also helps to solve another related problem.
In the simple picture mentioned above, starting with a black hole that contains $N$ qubits and adding to it another $N$ qubits from the infalling particles, we end up with a final black hole state that contains $2N$ qubits which certainly does not look like the vacuum. In our picture with negative qubits,
at the end of black hole evaporation, we end up with $N$ positive and $N$ negative qubits that have vanishing total energy and entropy. This final vacuum (or black hole) state has a qubit structure which is identical to that of the Unruh vacuum that describes the initial near horizon vacuum.

It is well--known that unitary evolution of black holes implies that the entanglement entropy of Hawking radiation should follow the Page curve[\PAG]. However, based on the simple picture of entanglement between the black hole and Hawking radiation above, $S_{ent}$ increases linearly during black hole evaporation until it reaches a maximum. This is not surprising since this analysis does not take into account how the negative qubits interact with the positive qubits that describe the black hole. 
In our model, unitarity is preserved due to the formation of positive--negative qubit pairs and a projection on the black hole state
at the end of each computation after the Page time. The projection eliminates the entanglement between the black hole and the radiation resulting in a pure final state for Hawking radiation. The same projections decrease $S_{ent}$
and reproduce the Page curve.

Even if black hole evaporation is unitary, the mechanism through which information comes out of the black hole remains an open question. Consider a shell of particles imploding to form the black hole in a pure but maximally entangled state. This
basically means the black hole can be described by a product of Bell states for any division into to equal subsystems. This state is completely scrambled and information is stored nonlocally among the qubits on the horizon. It is usually assumed that information comes out of the black hole in late (i.e. after Page time) radiation. 
As we describe below, in our model, information comes out of the black hole by teleportation. The simplest possibility is to teleport
the information in black hole Bell states directly. However, as we show in the Appendix this is not possible since a criterion for choosing the correct projection does not exist. Therefore, we need an alternative mechanism for teleportation. We find that  
in order to teleport the information we first have to describe the black hole Bell states in terms of products of two qubits. This is
achieved by a fine tuned interaction between the qubits of the black hole Bell states which is the inverse of premeasurement.
The entangled pairs of particles created near the horizon act as the mediators for the teleportation of information to Hawking radiation. 

As stated above, we assume that black holes are quantum computers and describe their unitary evaporation by the operations of a quantum computer. The initial state of the black hole is described by $N=S_{bh}/ln 2$ positive qubits. The state of the near horizon region, the Unruh vacuum, is described by $N$ positive--negative qubit pairs which carry zero energy and entropy.
The black hole qubits are in a pure but maximally entangled state given by $N/2$ Bell states whereas the two qubit pairs are in the Unruh state. This constitutes the input to the quantum computer with a total of $3N$ qubits.
We then create maximally entangled pairs of qubits (i.e. Bell states) near the horizon by the application of Hadamard and CNOT gates on the positive--negative qubit pairs of the Unruh state.
The positive qubits describe Hawking particles whereas negative ones are the infalling particles. After the emission of $n$ positive qubits, $S_{ent} = n ln 2$ and $S_{bh} = (N-n)ln2$.

Around Page time when half of the qubits are emitted so that $S_{ent}= S_{bh}= (N/2) ln2$, in addition to the emission of positive qubits, the quantum computer starts acting on the internal black hole states. First, an interaction between the qubits of a black hole Bell state, $H_{int}$, turns on for a fine tuned period of time. The unitary evolution due to this interaction, given by $U=e^{iH_{int}t_0}$, transforms the Bell state into a product of two qubits.
This is necessary in order to teleport one qubit of information at a time out of the black hole. The maximally entangled Hawking pair of qubits act as mediators for teleportation.
Each infalling negative qubit pairs with a positive black hole qubit to form a two qubit state.
Then, this two qubit state undergoes a CNOT and Hadamard gate which results in a state in which the internal black hole qubits are maximally entangled with the outgoing Hawking qubits. Finally, there is a projection on the black hole Hilbert space that eliminates the entanglement between the black hole and Hawking radiation. As a result, the final state of Hawking radiation is pure and black hole evaporation is unitary.
The same procedure also teleports the information in the positive black hole qubit to the outgoing radiation. After Page time, each emission of a Hawking qubit is accompanied by two of the above procedures which overall results in the decrease of $S_{ent}$ by $ln2$ at each step reproducing the Page curve. 

Black hole evaporation is obtained by repeating the above procedure $N$ times in series.
After the emission of $N$ positive qubits from the black hole we are left with a vacuum state with $N$ positive--negative qubit pairs and the the Hawking radiation with $N$ positive qubits in a pure state. 
The qubit structures of the Hawking radiation and final vacuum are identical to those of the initial black hole state and initial Unruh vacuum respectively. The information in the final state of the radiation can only be read by someone who knows the 
physics of the black hole, i.e. the operation $U=e^{iH_{int}t_0}$ since the original black hole Bell states can only be recovered by acting with $U^{\dagger}$ on the radiation. 

Our model is a concrete realization of the black hole final state idea of ref. [\HM] in terms of quantum computations with some differences. First, instead of imposing a final black hole state at the singularity we impose a projection at the end of each computation (in series). When the black hole completely evaporates we end up with a vacuum state that is very similar to the final state of ref. [\HM]. 
Second, our model has a well--defined and simple S--matrix, $S=U^{\dagger}=e^{-iH_{int}t_0}$ whereas in ref. [\HM] this S--matrix is assumed to be random and nonlocal in terms of the qubits. 
Finally and perhaps most importantly, the final state idea of ref. [\HM] suffers from problems with causality, i.e. time can run both forward and backward. In our model, there is an arrow of time since the projection on the black hole Hilbert space is not unitary and thus the whole process is not time reversible. 

This paper is organized as follows. In the next section, we describe our model for unitary black hole evaporation. It contains our assumptions, the input, the computation, and the output of the quantum computation. 
In section 3, we show that our model is a concrete realization of the black hole final state idea in terms of quantum computation. Section 4 contains a discussion of our results and our conclusions.


\bigskip
\centerline{\bf 2. A Black Hole Model with Unitary Hawking Radiation} 

In this section we describe a model of Hawking radiation that results in unitary black hole evaporation with no information loss.
Any such description must include some new and unconventional ideas that go beyond semiclassical gravity and quantum mechanics. 
Our model is based on the following assumptions.

1. Black holes are (modified) quantum computers. 
Therefore, we should be able to describe black hole evaporation in terms of the operations of a quantum computer. Such a quantum computer should take in an input which corresponds to the initial state of the black hole and its near horizon region. At the end of the computation it should give out an output that describes the final states of the black hole and the Hawking radiation. The transition between the input and output should be described by operations or gates of the quantum computer which are repeated in series $N$ times. At all stages, states of the system must be described in terms of qubits on which the operators act. 

2. There are two types of qubits. The first type,``positive" qubits carry positive energy of $1/R_{bh}$ and entropy of $ln2$. These are the usual qubits that describe the black hole. The second type,``negative' qubits have negative energy of $-1/R_{bh}$ and entropy of $-ln2$ (similar to infalling particles). It is well--known that inside the horizon energy is space--like and therefore can take both positive and negative values. Then, the first law of thermodynamics implies that qubits with negative energy must also have negative entropy. This seems to be allowed by semiclassical gravity and thermodynamics.
We also assume that a positive and a negative qubit can form a two qubit state with vanishing energy and entropy. The Unruh vacuum state in the near horizon region and the final state of the black hole are described by $N$ such pairs.

3. In the black hole, the scrambled information is stored in (the maximally entangled) Bell states. We assume that after Page time unitary interactions between the qubits of a Bell state turn on for a fine tuned amount of time. As a result, the Bell state turns into a product of two qubits. This process is exactly the inverse of the premeasurement[\PRE] that is well
understood in quantum measurement theory. It is these qubits of the product state that are teleported out of the black hole and constitute Hawking radiation.

4. As long as black hole evaporation is unitary there may be unobservable violations of quantum mechanics inside the black hole. Thus, the black hole is not exactly a quantum computer. In addition to unitary operations on the qubits, it also applies a nonunitary
projection on the internal black hole qubits. This projection fixes the final black hole state, disentangles Hawking radiation from the black hole and teleports the information to Hawking radiation.
It is similar to the fixed final black hole state imposed at the singularity in ref. [\HM] even though in our model the projection is 
made on the horizon at the end of each of the $N$ computations in series.

\medskip
{\noindent \bf 2.1 The Unitarity Problem}
\medskip

As mentioned in the introduction, the simplest version of the unitarity problem is usually described as follows[\MAT]. Assume that initially the black hole is in a pure state. Consider the creation of a pair of maximally entangled particles near the black hole horizon, e.g. in a Bell state.
One of the particles escapes to infinity to become part of the Hawking radiation whereas the other one falls into the black hole. Semiclassically, during this process, the black hole mass and entropy decrease by $1/R_{bh}$ and $ln2$ respectively. Because the created pair is entangled, each pair creation increases the entanglement entropy of the radiation (or the black hole), $S_{ent}$, by $ln2$. At the end of black hole evaporation, after the emission
of $N \sim S_{bh}$ Hawking particles, the black hole completely disappears leaving only the Hawking radiation. Since the entanglement entropy
of the radiation is $S_{ent}= N ln2$ the final Hawking radiation is in a mixed state. As first noticed by Hawking, the semiclassical description of black hole evaporation seems to take a pure state into a mixed state, i.e. it is not unitary.

The above analysis completely ignores the states that make up the thermodynamic entropy of the black hole and how it decreases as the the black hole evaporates. Naively, the number of qubits in the black hole increases by one with every pair creation. Thus, the thermodynamic entropy of the black hole seems to increase rather than decrease by the absorption of the infalling qubits. Moreover, by the end of evaporation, the black hole seems to contain $2N$ qubits (with vanishing energy and entropy) which does not look like the vacuum. This is a hint about the existence of a black hole final state which must be described by $2N$ qubits rather than nothing.

A simple solution to these problems is to assume that the infalling particles have negative entropy in addition to negative energy, i.e. they are what we called negative qubits above. 
The absorption of negative qubits decreases $S_{bh}$ as required. Moreover, the final black hole state is made of $N$ positive and $N$ negative qubits with vanishing energy and entropy.
However, this raises a new problem. If $S_{bh}$ decreases by the absorption of negative qubits then after the absorption of $N/2$ such qubits we have $S_{bh}=S_{ent}$. On general grounds, entanglement entropy is bounded by the coarse grained entropy,
$S_{ent} \leq S_{bh}$. Since $S_{bh}$ decreases, $S_{ent}$
cannot continue to increase after this point and must start to decrease. This is basically the problem of reproducing the Page curve.
Naively, there is no mechanism for changing the behavior of $S_{ent}$ which seems to monotonically increase throughout black hole evaporation. In the remainder of this section, we show that there exist a number of operations (applied in series) the quantum computer can apply on the black hole and infalling qubits that result in unitary black hole evaporation. Moreover, during evaporation
information is preserved and $S_{ent}$ follows the Page curve. 


\medskip
{\noindent \bf 2.2 The Input}
\medskip

We begin by describing the input to the quantum computer, i.e. the initial state of the black hole and its near horizon region in terms of qubits.
We assume that initially the black hole is in a maximally entangled pure state. In this case, maximally entangled means that any two equal size subsystems of the black hole are maximally entangled.
A generic state of $N$ qubits can be written as 
$$\ket{\psi}=\sum_{q_i} c_{12 \ldots N} \ket{q_1q_2 \dots q_N} \quad, \eqno(1)$$
where the single qubits $\ket{q_i}$ take the values $\ket{0}$ or $\ket{1}$, the sum is over all possible $2^N$ permutations and
$c_{12 \ldots N}$
are complex coefficients. However, this is not a maximally entangled state. Such a state can be written in terms of $N/2$ Bell states of qubits that belong to (any of) its two subsystems A and B. 
$$\ket{\psi}_{bh}=\left({1 \over 2}\right)^{N/4} (\ket{0}_A \ket{0}_B- \ket{1}_A \ket{1}_B)^{\otimes N/2} \quad. \eqno(2)$$
Here, we assumed that all $N/2$ Bell states are the same singlet state but in general any combination of the four Bell states given by 
$$\ket{\beta_{00}}={1 \over {\sqrt{2}}}(\ket{0}_A \ket{0}_B+ \ket{1}_A \ket{1}_B) \quad, \eqno(3)$$
$$\ket{\beta_{10}}={1 \over {\sqrt{2}}}(\ket{0}_A \ket{0}_B- \ket{1}_A \ket{1}_B) \quad, \eqno(4)$$
$$\ket{\beta_{01}}={1 \over {\sqrt{2}}}(\ket{1}_A \ket{0}_B+ \ket{1}_A \ket{0}_B) \quad, \eqno(5)$$
$$\ket{\beta_{11}}={1 \over {\sqrt{2}}}(\ket{1}_A \ket{0}_B- \ket{1}_A \ket{0}_B) \quad, \eqno(6)$$
can appear in $\ket{\psi}_{bh}$.
Thus, the most general maximally entangled black hole state can be written as
$$\ket{\psi}_{bh}= (\ket{\beta_{00}})^{\otimes N_1} (\ket{\beta_{10}})^{\otimes N_2} (\ket{\beta_{01}})^{\otimes N_3} (\ket{\beta_{11}})^{\otimes N_4} \quad,  \eqno(7)$$
where $N_1+N_2+N_3+N_4=N/2$.

In addition to the black hole state in eq. (7), the input also describes the state of the vacuum in the near horizon region.
This component of the input carries no energy or entropy since it describes the vacuum. Therefore, it has to be made of 
positive--negative qubit pairs. It is well--known that this state is the Unruh vacuum.
Let us denote positive and negative qubits by $\ket{q_i}_{\pm}$. Then, the Unruh vacuum can be written as a maximally entangled state
$$\ket{\psi}_{vac}=2^{-N/2} \sum_j \ket{q_j}_- \ket{q_j}_+   \quad, \eqno(8)$$
where $j=1,2, \ldots,N$. (The usual Unruh vacuum includes a Boltzmann factor $e^{-\pi E_j}$ in the sum. Here we are using the vacuum in the microcanonical ensemble with fixed energy so this term is absent.)

The input is obtained by combining the black hole and vacuum states in eqs. (7) and (8) respectively. 
Then
$$\ket{\psi}_{in}=2^{-N/2} \left[(\ket{\beta_{00}})^{\otimes N_1} (\ket{\beta_{10}})^{\otimes N_2} (\ket{\beta_{01}})^{\otimes N_3} (\ket{\beta_{11}})^{\otimes N_4}\right] \otimes \left(\sum_k \ket{q_k}_- \ket{q_{k}}_+ \right) \quad. \eqno(9)$$
The first term describing the black hole is made of $N$ positive qubits and lives in a Hilbert space of dimension $2^N$. The second term describing the Unruh vacuum is made of $N$ positive--negative qubit pairs and lives in another Hilbert space of dimension
$2^{2N}$.

\medskip
{\noindent \bf 2.3 The Computation}
\medskip

The quantum computer now begins operating on the input in eq. (9). The computation consists of different operations on the qubits as described below. These operations are repeated $N$ times in series until the end of evaporation.


1. Pair creation near the horizon: Throughout evaporation, every $t \sim R_{bh}$ a pair is created near the horizon. Thus our first step is to create a pair of entangled particles near the horizon from the vacuum state. For concreteness, let us consider the two qubit state $\ket{0}_- \ket{0}_+$ in the Unruh vacuum.
We first act on the qubit $\ket{0}_-$ with the Hadamard gate[\NIE] defined by the operations
$$\ket{0} \to  {1 \over {\sqrt{2}}}(\ket{0}+ \ket{1}) \qquad \ket{1} \to  {1 \over {\sqrt{2}}}(\ket{0}- \ket{1}) \quad, \eqno(10)$$
and obtain the state  
$$\ket{0}_-\ket{0}_+ \to  {1 \over {\sqrt{2}}}(\ket{0}_-+ \ket{1}_-) \ket{0}_+ \quad. \eqno(11)$$ 
We then act on this state with the
CNOT gate[\NIE] which does not change the second qubit if the first one is $\ket{0}$ and switches it if the first qubit is $\ket{1}$. Thus, 
we end up with 
$$(\ket{0}_- \ket{0}_+) \to {1 \over {\sqrt{2}}}(\ket{0}_- \ket{0}_+  + \ket{1}_- \ket{1}_+)= \ket{\beta _{00}}_{\pm} \quad, \eqno(12)$$
which describes the Bell state created near the horizon. This is just the standard way to produce Bell states from product states. It is easy to see that the same procedure takes the other three input states into the other three Bell states, i.e.
$$(\ket{0}_- \ket{1}_+) \to \ket{\beta_{01}}_{\pm} \qquad (\ket{1}_- \ket{0}_+) \to \ket{\beta_{10}}_{\pm} \quad (\ket{1}_- \ket{1}_+) \to \ket{\beta_{11}}_{\pm} \quad. \eqno(13)$$ 
Thus, the form of the two qubit product in the Unruh vacuum determines the detailed form of the maximally entangled pair created near the horizon.
After the creation of one pair, the total state is given by eq. (9) with one qubit pair of the form $\ket{q_k}_- \ket{q_{k}}_+$ replaced by one $\ket{\beta_{ij}}_{\pm}$. We take this this to be the concrete case given in eq. (12); i.e. with $\ket{\beta_{00}}_{\pm}$.

2. Hawking radiation: The above pair creation process is repeated $N/2$ times until Page time, $t_P =N R_{bh}/2$. 
The negative qubits, $\ket{q}_-$, in the pair are the infalling qubits whereas the outgoing positive qubits, $\ket{q}_+$, are the Hawking particles. Until Page time $N/2$ maximally entangled pairs $\ket{\beta_{ij}}_{\pm}$ are created. Each emission of a positive qubit increases $S_{ent}$ by $ln 2$ whereas each absorption of a negative qubit reduces $S_{bh}$ by the same amount.
At Page time the qubits are divided as follows. There are $N/2$ outgoing Hawking and infalling qubits, $\ket{\beta_{ij}}_{\pm}$, which are maximally entangled. In addition there are
$N/2$ positive--negative qubit pairs in the Unruh vacuum and $N$ positive qubits, $\ket{\beta_{ij}}$, in the black hole. Then, total state is given by 
$$\ket{\psi}_{tot}=2^{-N/2}(\ket{\beta_{ij}})^{\otimes N/2}) \otimes (\ket{\beta_{ij}}_{\pm})^{\otimes N/2}) \otimes \left(\sum_n \ket{q_n}_- \ket{q_n}_+ \right) \quad, \eqno(14)$$
where $n=1,2 \ldots N/2$ and the three terms are shortcuts for the black hole state, the maximally entangled infalling and outgoing pairs and the Unruh vacuum respectively. 

This state has $S_{bh}=S_{ent}=(N/2) ln2$ with $S_{bh}$ ($S_{ent}$) monotonically decreasing (increasing). It is well--known that for a unitary theory, the smaller subsystem of an originally pure system is maximally entangled with the larger subsystem. This means that $S_{ent}$ equals the coarse grained entropy of the smaller subsystem. After Page time, the black hole becomes the smaller subsystem with decreasing $S_{bh}$ and thus $S_{ent}$ must decrease monotonically to zero. Reproducing the Page curve, with $S_{ent}$
monotonically increasing (decreasing) until (after) Page time is considered to be equivalent to proving unitarity of black hole evaporation. The steps below describe the operations that lead to unitary black hole evaporation with $S_{ent}$ reproducing the Page curve.

3. Interactions between qubits of the black hole Bell states: We can try to teleport the black hole Bell states directly to Hawking radiation but as we show in the Appendix, there is no criterion for picking a projection that preserves information. More concretely, when the black hole Bell state is teleported by using two Hawking pairs as
mediators the state becomes a superposition of 16 states. Only one of these contains the correct information that originally belongs to the black hole. Any projection on the black hole state eliminates the entanglement between the black hole and Hawking radiation and thus preserves unitarity. However, there seems to be no general criterion for picking the (only) correct projection that preserves information. Therefore, we conclude that directly teleporting Bell states to Hawking radiation is not useful for our purposes.

Therefore, we need a new mechanism for teleportation of information out of the black hole. As we show below, this can be achieved by transforming black hole Bell states into a product of two qubits.
We assume that every $t \sim R_{bh}$ after Page time, when an entangled pair is created on the horizon,
this is accompanied by the transformation of a black hole Bell pair into a product of two qubits. Just like the Bell states each of these qubits carries information about the black hole state. 
Every $t \sim R_{bh}$ the information in two such qubits will be teleported out of the black hole to the positive qubits which belong to early radiation. 

There is no unitary transformation of the form $U=U_A U_B$, that acts on each qubit in eqs. (3)-(6) separately, that can transform
a Bell state into a product state. However, we can achieve this if we assume that
there is an interaction between the qubits of the Bell state that lasts for a fine tuned period of time. This is basically the inverse of what is called premeasurement in quantum measurement theory[\PRE]. Since the premeasurement is unitary its inverse exists and is also unitary. Consider an interaction between the qubits of the black hole Bell state of the form 
$$H_{int}=\epsilon(1+\sigma_z^A)\sigma_x^B \quad, \eqno(15)$$
where $\epsilon<<1$ parametrizes the strength of the interaction.
The operator $\sigma_z$ acts on the qubits as $\sigma_z \ket{0}=\ket{0}$ and $\sigma_z \ket{1}=-\ket{1}$, i.e. $\ket{0},\ket{1}$ are the spin up and down in the z direction. On the other hand, $\sigma_x \ket{0}=\ket{1}$ and $\sigma_x \ket{1}=\ket{0}$
and so the eigenstates of $\sigma_x$ are 
$$\ket{0}_x={1 \over {\sqrt{2}}}(\ket{0}+\ket{1}) \qquad \ket{1}_x={1 \over {\sqrt{2}}}(\ket{0}-\ket{1}) \quad, \eqno(16)$$
with eigenvalues $\pm 1$ respectively.

Premeasurement starts with an initial product state, for example
$$\ket{\phi}=(a \ket{0}_A+b \ket{1}_A) \ket{1}_B \quad, \eqno(17)$$
where where $|a|^2+|b|^2=1$ and $\ket{1}_B=(1/\sqrt{2})(\ket{0}_x-\ket{1}_x)$.
Due to the interaction, after a time $t$ the state becomes
$$\ket{\phi(t)}=U(t)\ket{\phi}=e^{\epsilon t(1+\sigma_z^A)\sigma_x^B }[(a\ket{0}_A+b\ket{1}_A) \ket{1}_B)] \quad, \eqno(18)$$
which leads to
$$\ket{\phi(t)}={1 \over {\sqrt{2}}} \left(a e^{2it\epsilon} \ket{0}_A \ket{0}_x^B + b \ket{1}_A \ket{0}_x^B - a e^{-2it\epsilon} 
\ket{0}_A \ket{1}_x^B - b \ket{0}_A \ket{0}_x^B \right) \quad. \eqno(19)$$
Using the eigenvalues of the different states under $\sigma_x$ and $\sigma_z$ we obtain
$$\ket{\phi(t)}= \left(-ia sin(2t \epsilon) \ket{0}_A \ket{0}_B +b \ket{1}_A \ket{1}_B
+ a cos(2t \epsilon) \ket{0}_A \ket{1}_B \right) \quad. \eqno(20)$$
Now if the interaction acts for a period of $t=\pi/4 \epsilon$ the state above becomes
$$\ket{\phi(\pi/4 \epsilon)}=-i(a \ket{0}_A \ket{0}_B +ib \ket{1}_A \ket{1}_B) \quad, \eqno(21)$$
which is a Bell state. Thus, we find that the unitary interaction in eq. (15) transforms the product state in eq. (17) into a Bell state.
Since the evolution by $U$ is unitary, its inverse that takes a Bell state to a product of qubits is given by $U^{\dagger}$. From eq. (21) we see that under $U^{\dagger}(\pi/4 \epsilon)$ the black hole Bell state becomes 
$$U^{\dagger}(\pi/4 \epsilon) (\ket{0}_A \ket{0}_B- \ket{1}_A \ket{1}_B) =  (i \ket{0}_A- \ket{1}_A) \ket{1}_B \quad. \eqno(22)$$ 
It is easy to check that the other three Bell states also become products of two qubits of the form above. Therefore, under the action of $U^{\dagger}(\pi/4 \epsilon)$ a generic black hole Bell state turns into a product of two qubits as 
$$\ket{\beta_{ij}}=(a \ket{0}_A+b \ket{1}_A) (c \ket{0}_B+d \ket{1}_B) \quad, \eqno(23)$$
with the coefficients (a,b)(c,d) corresponding to different Bell states
$$\ket{\beta_{10}}:(i/\sqrt{2},-1/\sqrt{2})(0,1) \qquad \ket{\beta_{00}}:(i/\sqrt{2},1/\sqrt{2})(0,1) \quad, \eqno(24)$$
and
$$\ket{\beta_{01}}:(i/\sqrt{2},-1/\sqrt{2})(1,0) \qquad \ket{\beta_{11}}:(i/\sqrt{2},1/\sqrt{2})(1,0) \quad. \eqno(25)$$
To summarize, after the above interaction, each black hole Bell state becomes a product of two qubits: one of the form 
$(i \ket{0} \pm \ket{1})/\sqrt{2}$ and the other either $\ket{0}$ or $\ket{1}$.

4. Unitary black hole evaporation and teleportation of information:
After their creation, each black hole qubit of the form $a \ket{0}_{A,B}+b \ket{1}_{A,B}$ will be coupled to a negative infalling qubit which is entangled with an outgoing positive qubit, i.e. early Hawking radiation. Each of the operations below occur twice every $t \sim R_{bh}$ once for each qubit in the product in eq. (23).

The state that describes one black hole qubit coupled to the entangled infalling--outgoing pair is (we suppress the black hole index $A,B$ from now on)
$$\ket{\psi}={1 \over 2} (a \ket{0}+b \ket{1}) (\ket{0}_-\ket{0}_+  + \ket{1}_- \ket{1}_+)  \quad. \eqno(26)$$
Expanding, we get 
$$\ket{\psi}={1 \over 2} [a \ket{0}(\ket{0}_-\ket{0}_+  + \ket{1}_- \ket{1}_+) + b \ket{1} (\ket{0}_-\ket{0}_+  
+ \ket{1}_- \ket{1}_+)] \quad. \eqno(27)$$

The black hole part of eq. (27) can have any of the four choices for $a,b$ in eqs. (24)-(25). The part that describes the entangled Hawking pair can also be any of the four Bell states. Thus, overall there are sixteen states of the form given in eq. (27) and they all lead to similar results when acted upon by the operations below.

We describe the absorption of the infalling qubit by the black hole by coupling the qubits $\ket{q}$ and $\ket{q}_-$ and considering them as a pair. Thus, the state in eq. (27) describes the emission of the qubit $\ket{q}_+$ from the black hole.
This process decreases the black hole energy and entropy by $1/R_{bh}$ and $\ln 2$ respectively since the black hole absorbs 
the negative qubit $\ket{q}_-$. The entanglement entropy, $S_{ent}$, of the black hole (or Hawking radiation) increases by $\ln 2$ due to the entanglement between $\ket{q}_-$ and $\ket{q}_+$.

As mentioned above, after Page time, $S_{ent}$ has to decrease. If we can find a procedure to achieve this, then our model will reproduce the Page curve. Below, we will describe a number of operations that act on the state in eq. (27) and decrease $S_{ent}$ by 
$ln 2$ at every step, i.e. every $t \sim R_{bh}$. 

We now act on the internal pair of qubits ($\ket{q} \ket{q}_- $) in eq. (27)  with the CNOT gate and obtain
$$\eqalignno{
\ket{\psi}={1 \over 2} [a \ket{0}(\ket{0}_-\ket{0}_+  &+ \ket{1}_- \ket{1}_+) &(28)  \cr
&+  b \ket{1} (\ket{1}_-\ket{0}_+  + \ket{0}_- \ket{1}_+) } $$
Next we act with the Hadamard gate on the black hole qubit $\ket{q}$ which gives
$$\eqalignno{
\ket{\psi}={1 \over 2} [a(\ket{0}+ \ket{1}) (& \ket{0}_-\ket{0}_+  + \ket{1}_- \ket{1}_+) &(29) \cr
&+  b(\ket{0}- \ket{1}) (\ket{1}_-\ket{0}_+  + \ket{0}_- \ket{1}_+)] } $$
This can now be rewritten as
$$\eqalignno{ 
\ket{\psi}={1 \over 2} [& \ket{0} \ket{0}_- (a \ket{0}_+ +b \ket{1}_+)+  \ket{0} \ket{1}_- (a \ket{1}_+ + b\ket{0}_+) &(30) \cr
&+  \ket{1} \ket{0}_- (a\ket{0}_+-b \ket{1}_+)+  \ket{1} \ket{1}_- (a \ket{1}_+-b \ket{0}_+)]  } $$
In the final state given by eq. (30) we observe that the internal state of qubits ($\ket{q} \ket{q}_- $) are maximally entangled with the outgoing qubit $\ket{q}_+$. At this stage Hawking radiation is maximally entangled with the black hole and Hawking radiation does not contain any information. 

Finally, we act with the projection operator $P=(\ket{0}\ket{0}_{- -} \bra{0}\bra{0})$ on the black hole state in eq. (30).
This projection reduces the entangled state in eq. (30) to the product state
$$\ket{\psi}={1 \over 2} (\ket{0} \ket{0}_-)(a \ket{0}_+ + b \ket{1}_+) \quad, \eqno(31)$$
The projection achieves three aims. First, it destroys the entanglement between the internal black hole state and Hawking radiation. Therefore, it decreases $S_{ent}$ by $ln 2$.
Second, the projection teleports the information in the black hole state to Hawking radiation. Note that the state of the outgoing qubits in eq. (31) is exactly the initial black hole state in eq. (26).
This is nothing but the usual teleportation[\TEL] of the black hole qubit using the entangled pair of qubits created on the horizon as mediators. However, at the end rather than making a measurement of the internal qubits and sending this information to
the outgoing qubit we used a projection operator. Third, the projection determines the (component of the) unique black hole final state given by
the two qubit state $\ket{q} \ket{q}_-$. In eq. (31), this is $\ket{0} \ket{0}_-$.

From eq. (30) it is clear that there are other possible projections on the internal state which lead to other states for the outgoing qubit. These will cause the wrong information to be teleported. The correct projection is determined by the Unruh vacuum qubits in the input given by eq. (8). In our case these were $\ket{0}_- \ket{0}_+$ thus the projection operator is 
$P=(\ket{0} \ket{0}_{- -}\bra{0} \bra{0})$. Another initial vacuum state will lead to a different projection which still teleports the correct information to radiation.
For example. consider the case in which the Unruh vacuum qubits are $\ket{1}_- \ket{0}_+$ which after applying the Hadamard and CNOT gates become the entangled qubit pair $\ket{\beta_{10}}=(\ket{0}_- \ket{0}_+ - \ket{1}_- \ket{1}_+)$. Going through the same steps above we obtain the final state
$$\eqalignno{
\ket{\psi}={1 \over 2} &[ \ket{0} \ket{0}_- (a \ket{0} -b \ket{1})+  \ket{0} \ket{1}_- (-a \ket{1} +b \ket{0}) &(32) \cr
&+  \ket{1} \ket{0}_- (a \ket{0} +b \ket{1})+ \ket{1} \ket{1}_- (-a \ket{1} -b \ket{0}) }$$
We now see that in order to teleport the correct black hole information we need to project with 
$P=(\ket{1} \ket{0}_{- -}\bra{0} \bra{1})$. Again this is determined by the vacuum qubits we started from, i.e. 
$\ket{1}_-\ket{0}_+$. 
The initial vacuum state not only determines the correct projection that disentangles the Hawking radiation and teleports the information but it also fixes the final black hole qubit state. From the above two examples, it is clear that when
the Unruh vacuum qubits are $\ket{0}_- \ket{0}_+$ (or $\ket{1}_- \ket{0}_+$) the qubits that describe (this component of the) final black hole state are the same, i.e. $\ket{0} \ket{0}_-$
(or $\ket{1} \ket{0}_-$). Surprisingly, we find that there is  a lot of information contained in the exact qubit structure of the 
initial Unruh vacuum.
When the above procedure for disentanglement occurs twice, once for each black hole product qubit, $S_{ent}$ decreases overall by 
$ln 2$ (taking into account the absorption of $\ket{q}_-$ which increases $S_{ent}$ by $ln 2$). This is precisely what is required to obtain the Page curve; $S_{ent}$ decreases by $ln2$ with each emission, every $t \sim R_{bh}$ after Page time.

Since at Page time there are $N/2$ negative qubits in the black hole and $N/2$ entangled pairs to be produced from the vacuum, the steps above continue until $t = N R_{bh}$. The vacuum state left after the black hole completely evaporates is given by
$N$ positive--negative qubit pairs the type $\ket{q} \ket{q}_-$, i.e. $\sum (\ket{q} \ket{q}_-)$.
As expected, this state carries no energy or entropy since it is made of pairs of positive and negative qubits. Its qubit structure is identical to that of the initial Unruh vacuum.

We pushed our description of the black hole evaporation until the very end. However, since our model requires negative qubits and these exist due to the metric inside the horizon, our model actually breaks down when the black hole metric can no longer be trusted. Assuming that this happens when the black hole radius is $O(10)$ times the Planck length we find that our model fails when $S_{bh}$ or the number of positive black hole qubits is around $O(10^2)$. After this point, the physics can only be described by the correct theory of quantum gravity. With this caveat, for simplicity below we describe the final state or output by pretending that our model 
works until the very end of evaporation. 

As mentioned above, the nonunitary projection on the final black hole state in eq. (30) is what preserves the unitarity of black hole evaporation. Without this projection, Hawking radiation remains maximally entangled with the black hole and information does not come out of the black hole. In that case, at the end of evaporation, Hawking radiation is in a mixed state with only $1/4$ of the 
initial information released into the radiation. The other $3/4$ of the information cannot be emitted and are lost forever.

\medskip
{\noindent \bf 2.4 The Output}
\medskip

After Page time, the evaporation process continues with pair creations every $t \sim R_{bh}$ for an additional $N/2$ times. Each pair creation and emission of a positive qubit is accompanied by two disentanglement processes of the type above. Thus, overall
$N$ positive qubits are emitted as Hawking radiation.
At time $ t=N R_{bh}$ all the the initial near horizon qubits are exhausted and all the information in the positive black hole qubits is teleported to the Hawking radiation. We are left with the output which consists of the vacuum state left after the black hole evaporates and Hawking radiation given by the state
$$\ket{\psi}_{out}=(2)^{-N/2}\left(\sum_n \ket{q_n} \ket{q_n}_- \right) \otimes [(a_i \ket{q_i}_+ +b_i \ket{q_i}_+)]^{\otimes N} \quad. \eqno(33)$$
Since this is a product state, Hawking radiation is in a pure state which solves the unitarity problem.
In addition, all the initial information in the black hole, which consists of $a_i,b_i$, comes out in the Hawking radiation. The information is transferred from the black hole to Hawking radiation by teleportation. This solves the information problem.

We started with the input in eq. (9) which consists of the initial black hole state and that of the Unruh vacuum. The output in eq. (33) describes the final black hole state and Hawking radiation. We observe that, the qubit structure of Hawking radiation is the same as that of the initial black hole state. In fact, that is the definition information preservation in black hole evaporation.
In addition, the vacuum state left after the black hole evaporates is identical to
the initial Unruh vacuum state. It seems that, during black hole evaporation, the black hole and Hawking radiation qubits are simply exchanged. The same is true for the initial Unruh vacuum and the vacuum left after the black hole evaporates. This is a new and interesting way to look at Hawking radiation.

The final state of Hawking radiation contains the information in a way that is very difficult to decipher. Hawking radiation given by eq. (33) actually contains $N/2$ qubits of the form $i \ket{0}_+ \pm \ket{1}_+$ and $N/2$ qubits in either $\ket{0}_+$ or $\ket{1}_+$
states. These are not the Bell states for which the maximal entanglement of the Hawking radiation is manifest. 
Only an observer who knows black hole physics, i.e. the correct form of the product qubits, the interaction in eq. (15) between the qubits and the length of time it acts
can reconstitute the original Bell states that made up the initial black hole state given by eq. (9). More specifically, the observer
has to know the operator $U^{\dagger}(\pi/4 \epsilon)= e^{(\pi/4)(1+\sigma_z^A)\sigma_x^B}$ and act with it on a product of Hawking qubits of the type $(i \ket{0} \pm \ket{1}) \ket{0}$ or $(i \ket{0} \pm \ket{1}) \ket{1}$. This can be considered as the distillation of the information in Hawking radiation in terms of Bell states.
If the observer acts with $U^{\dagger}$ on other products of qubits she will not obtain Bell states or the information in the initial black hole state. Note that, when the Bell states are reconstructed there are 
numerous ways to get the number of each type of Bell state ($N_1,N_2,N_3,N_4$ in eq. (9)) wrong. However, all these states can be transformed into the correct state by unitary transformations of the form $U_AU_B$ and thus contain the same information.

It is surprising to find that unitary black hole evaporation is a relatively simple process, in the sense that it requires only three different operations, namely the CNOT and Hadamard gates, the unitarity violating projection,
and an interaction between qubits of the black hole Bell states. Ideally, we would like to have a computation that uses only unitary operations of a quantum computer. In this paper, we argue that this is not possible and unitarity of black hole evaporation ironically requires a nonunitary projection in the black hole. As stated above, 
we assume that this is not a problem as long as nonunitarity is unobservable.

\bigskip
\centerline{\bf 3. Relation of the Model to the Black Hole with a Final State Idea}
\medskip

Our model is a concrete realization of the black hole final state idea[\HM] in terms of a quantum computation with some differences. We first briefly review the results of ref. [\HM] and then show its relation to our model. Before evaporation the black hole Hilbert space can be written in product form as
${\cal H}={\cal H}_{bh} \otimes {\cal H}_A \otimes {\cal H}_B$ where $A,B$ denote the infalling and outgoing qubits respectively.
Each of the individual Hilbert spaces has dimension $D=2^{S_{bh}}$. 
The initial state before evaporation begins is given by
$$\ket{\psi}_{i}=\ket{\psi}_{bh} \otimes {1 \over {\sqrt{D}}} \sum_a \ket{a}_A \ket{a}_B \quad, \eqno(34)$$
where the infalling and outgoing qubits are maximally entangled. The idea of ref. [\HM] is to impose a black hole final state given by the density matrix 
$$\rho_{f}=\ket{\xi}\bra{\xi}_{bh,a} \otimes {{I_B} \over D} \quad. \eqno(35)$$
The state that couples the black hole to the infalling qubits and imposes the final state projection is
$$\ket{\xi}_{bh,a}={1 \over {\sqrt{D}}} \sum_a S^{\dagger} \ket{a}_{bh} \ket{a}_A \quad, \eqno(36)$$
where $S$ is a random and nonlocal unitary transformation.
The probability, $P(n)$ for any collection of measurements on the outgoing qubits can be expressed in terms of projections, 
$C=\Pi_1 \ldots \Pi_n$ as
$$P(n)={{Tr(\rho_f C^{\dagger} \rho_i C)} \over {\Sigma Tr(\rho_f C^{\dagger} \rho_i C)}}=\bra{\psi}S^{\dagger}C C^{\dagger} S \ket{\psi} \quad. \eqno(37)$$
We see that the final states of Hawking radiation is given by $S \ket {\psi}_B$ for any measurement. This is a pure state which solves the unitarity problem. Moreover, $S \ket {\psi}$ contains all the information in the initial black hole state $\ket {\psi}_{bh}$ which
solves the information problem. Information has been transferred from the black hole to Hawking radiation by teleportation (plus a
unitary transformation $S$).

It is clear that the above description is very similar to our model with some differences. 
Consider the initial state in eq. (34). This is the same as the input in our model
once we identify $\ket{\psi}_{bh}$ with the initial black hole state in eq. (8). The imposition of the black hole final state in eq. (35) is the sum of all the projections on black hole state at the end of each computation in our model. The unitary operation $S$ is nothing but the inter--qubit interaction we used in our model, i.e. $S=U^{\dagger}(\pi/4 \epsilon)=
e^{i \pi H/4 \epsilon}$ with the interaction Hamiltonian given by eq. (15). $S^{\dagger} \ket{a}_{bh}$ are the product qubits that we defined in eqs. (24)-(25). The final state of Hawking radiation 
$S \ket{\psi}_B$ is exactly the one we obtained in our model in eq. (33) for the outgoing qubits. In order to read the information it contains we need to know the precise form of $S=U^{\dagger}(\pi/4 \epsilon)$.

On the other hand, there are some differences between ref. [\HM] and our model.
First, the description in ref. [\HM] is in terms of quantum mechanical operations on the whole Hilbert space whereas in our model we have a large number of operations of a quantum computer that act in series on a few qubits at a time. Second, ref. [\HM] imposes a black hole final state at the singularity, i.e. at the end of evaporation whereas, in our model, the projections are made during evaporation around the times $t \sim (NR_{bh}/2)+ n R_{bh}$ where $n=1, \ldots, N/2$. 
Third,, in ref. [\HM] the S--matrix in eq. (36) is assumed to be random and nonlocal. In our model, $S$ is neither nonlocal nor random; it is simply $S=U^{\dagger}(\pi/4 \epsilon)=e^{(\pi/4)(1+\sigma_z^A)\sigma_x^B}$. 
Finally and perhaps most importantly, unlike the final state idea of ref. [\HM], our model does not suffer from problems of causality, 
In ref. [\HM], time may flow forward or backward[\TIM]. 
In our model, the arrow of time is fixed by the projections on the black hole state at the end of each computation. 
These projections are not unitary and thus not time reversible. Information is lost during the projections and cannot be recovered when we run time backwards. 

Since our model realizes the ideas of ref. [\HM], it also suffers from some of the same problems. For example, in this paper,
we completely neglected possible interactions between the black hole and infalling qubits. Such interactions inside the black hole 
are expected to affect the physics of Hawking radiation. However, in refs. [\GOT]
it was shown that these effects are not too large and the fidelity of the final Hawking radiation is relatively close to one. Based on these results, we assume that our description remains basically correct even in the presence of interactions between the black hole and the infalling qubits. Nevertheless, this is an important issue that needs to be further investigated in detail.

\bigskip
\centerline{\bf 4. Conclusions and Discussion}
\medskip

In this paper, we described unitary black hole evaporation with no information loss in terms of a quantum computation. This provides additional evidence that black holes can be described as quantum computers. Unitarity results from a (nonunitary) projection on the black hole state whereas information is preserved due to teleportation from the black hole to Hawking radiation. It is surprising that unitary black hole evaporation requires a nonunitary operation in the black hole just like in ref. [\HM]. We assume that this is allowed as long as nonunitarity is not observable outside the black hole as, for example, in the semiclassical Hawking radiation.
In any case, it is important to find out whether unitary black hole evaporation can be described by only unitary operations or not due to a fundamental physical reason.
Another surprise is the fact that unitary evaporation is a relatively simple process; it requires
only four operations, namely the interaction between the black hole Bell states, the Hadamard and CNOT gates and the projection on the black hole state.

The details of black hole evaporation determine the existence of firewalls[\FIRE]. According to accepted wisdom,
if black hole evaporation is unitary, then late radiation has to be entangled with early radiation in order to purify it. However, the outgoing qubits also have to be maximally entangled with the infalling ones for the horizon to be smooth. This is not allowed by the monogamy of entanglement. Preserving unitarity forces us to give up smooth horizons; i.e. there are firewalls.
On the other hand, if black holes evaporate according to our model, then horizons are smooth and there are no firewalls. 
In our model, the pairs of qubits created near the horizon are maximally entangled leading to a smooth horizon. Much later, when the outgoing qubits are very
far from the horizon, they are disentangled from the black hole and the infalling qubits by the projection. At no time, do the outgoing qubits need to be maximally entangled with the early radiation to preserve unitarity. There is no problem with monogamy of entanglement. It is the projection on the black hole state that leads to unitarity rather than the entanglement between early and late radiation.

As mentioned above, if we compare the input in eq. (9) and the output in eq. (33), we find that, up to the unitary transformation
$S=U^{\dagger}(\pi/4 \epsilon)$ applied to the final Hawking radiation,
black hole evaporation can be described as an exchange of qubits. The $N$ positive qubits of the initial black hole state become those of the final Hawking radiation whereas the $N$ positive--negative qubit pairs describing the initial Unruh vacuum simply become those of the final vacuum left after the black hole evaporates completely. This raises the possibility of describing Hawking radiation as a exchange between the black hole and radiation qubits[\EDI]. In the framework of the ER=EPR correspondence[\ER] it is
possible for the black hole and radiation qubits to be exchanged through microscopic wormholes that connect them. In this case, black hole evaporation proceeds not in the visible space but through the microscopic wormholes.

At this stage, a quantum computational description of black hole evaporation by its very nature cannot describe the black hole spacetime or the physical laws that govern it. All we can say is that black hole physics can be reproduced by quantum computations
with ad hoc rules that we know must hold if this description has any merit.
For example, in section 2, we described the creation of entangled pairs from the Unruh vacuum by the
CNOT and Hadamard gates. In this description, the notion of a horizon or the physica that leads to pair creation i.e. quantum field theory near the horizon are completely obscure. The question is whether there are purely quantum computational concepts and principles that correspond to those of semiclassical or quantum gravity. For example, the existence of an entangled qubit structure like that of the Unruh vacuum may correspond to the existence of a horizon where the horizon area is given by the number of qubits $N$ (in Planck units). Alternatively simply the existence of negative qubits may imply the existence of a horizon. Similarly, in the quantum computation there is no notion of motion; therefore it is not clear if any qubits are infalling or outgoing. Of course, we can assign positive (negative) qubits to outgoing (infalling) particles but this is based purely on our knowledge of the spacetime picture and it cannot be inferred from the quantum computation. It is interesting to note that the existance of the Page curve for $S_{ent}$ can be motivated purely from a quantum computational; point of view since $S_{ent} \leq S_{bh}$ is a fundamental property of entanglement entropy. Actually, the Page curve is quite obscure in the spacetime picture which is basically equivalent to the information paradox.

Recently, the Page curve for the entanglement entropy of Hawking radiation (or a black hole) was obtained from the fine grained gravitational entropy in terms of the minimum contribution that arises from quantum extremal surfaces[\ALM]. It turns out that, the fine grained gravitational entropy has two separate contributions that arise from two different extremal surfaces. During black hole evaporation, the contribution of one of these is increases whereas that of the second one decreases exactly reproducing the Page curve. It would be interesting to find the connection between these results and our model. Even if it is not the model described in this paper, we expect that there is a quantum computational description of the results in ref. [\ALM].


\bigskip
\centerline{\bf Appendix}
\medskip

In this Appendix we show that one can teleport a black hole Bell state by using two Hawking Bell states (made of the infalling and outgoing qubits), one for each qubit of the black hole Bell state. However, we argue that there is no criterion to select the correct projection on the black hole state that preserves information. Therefore, we conclude that teleporting black hole Bell states is not useful for our purposes. We begin with the product state
$$(\alpha \ket{0}+ \beta \ket{1})_{12} (\ket{00}+ \ket{11})_{34}) (\ket{01}+ \ket{10})_{56})  \quad, \eqno(A.1)$$
where the qubits $1,2$ belong to the black hole. The qubits $3,4$ and $5,6$ and the two Hawking pairs where qubits $4,6$ are the outgoing Hawking particles. For Bell states 
$\alpha,\beta=\pm 1$ but we include the phases $\alpha, \beta$ in order to follow the information contained in them.
We arbitrarily picked the particular Bell states for the Hawking pairs but it is easy to show that our results hold for any other choice. We now would like to teleport the information in qubit $2$ to
qubit $4$ by using the Bell state $(\ket{00}+ \ket{11})_{34}$ and teleport the information in qubit $1$ to qubit $6$ by using 
$(\ket{01}+ \ket{10})_{56}$. For the former we apply the CNOT and Hadamard gates on qubits $2,4$ and $2$ respectively.
For the latter we apply the CNOT and Hadamard gates on qubits $1,6$ and $1$ 
respectively. These operations take the state in eq. (A.1) to
$$\eqalignno{
&(\alpha \ket{11} + \beta \ket{00})_{46} (\ket{0000}-\ket{1100})_{1235} + (\alpha \ket{11} - \beta \ket{00})_{46} (\ket{1000}-\ket{0100})_{1235}  \cr
&(\beta \ket{11} + \alpha \ket{00})_{46} (\ket{0011}-\ket{1111})_{1235} + (\beta \ket{11} - \alpha \ket{00})_{46} (\ket{0111}-\ket{1011})_{1235}  \cr
&(\alpha \ket{01} + \beta \ket{10})_{46} (\ket{0010}-\ket{1110})_{1235} + (\alpha \ket{01} - \beta \ket{10})_{46} (\ket{1010}-\ket{0110})_{1235}  \cr
&(\beta \ket{01} + \alpha \ket{10})_{46} (\ket{0001}-\ket{1101})_{1235} + (\beta \ket{01} - \alpha \ket{10})_{46} (\ket{0101}-\ket{1001})_{1235}  }$$
here the outgoing Hawking qubits $4,6$ are in maximally entangled Bell states. These are in turn entangled with the internal black hole qubits; these are the original black hole qubits $1,2$ and the infalling qubits $3,5$. In ordinary teleportation, the black hole state would be measured by an observer who would send that information classically to another observer who would manipulate the qubits $4,6$ properly (i.e. by using the operators $X$ and $Z$ on either qubit) to obtain the correct 
teleported state. In our case, the quantum computer needs to apply a projection to the black hole state, i.e. qubits $1,2,3,5$ that will pick the correct state to be teleported. This is the first state in the second line above, i.e. $(\beta \ket{11} + \alpha \ket{00})_{46}$, 
so the projection in this particular example has to be 
$(\ket{0011}-\ket{1111})_{1235} (\bra{0011}-\bra{1111})_{1235}$. There are 64 different states of the form in eq. (A.1) and for each one there is a different correct projection on the black hole state. Unfortunately, it seems that there is no general rule or criterion that determines the correct projection. Any projection destroys the entanglement between qubits $3,5$ and the rest and
preserves unitarity. However, only the correct projection preserves information.
Therefore, we conclude that the standard teleportation of black hole Bell states is not useful for our purposes.

\bigskip
\centerline{\bf Acknowledgments}

I would like to thank the Stanford Institute for Theoretical Physics for hospitality.

\vfill

\refout

\end
\bye